\documentclass[final,times]{elsarticle}
\usepackage{amssymb,amsmath}
\usepackage{srcltx}
\usepackage{xspace}
\usepackage{graphicx}

\newcommand{\grad}{\ensuremath{{}^{\circ}}\xspace}

\newcommand{\mumu}{\ensuremath{{\mu^{+}\mu^{-}}}\xspace}

\renewcommand{\epsilon}{\varepsilon}

\newcommand{\DD}{\ensuremath{D\overline{D}}\xspace}

\newcommand{\Gee}{\ensuremath{\Gamma_{ee}}\xspace}

\newcommand{\GBmumu}{\ensuremath{\Gamma_{ee}\times\Gamma_{\mu\mu}/\,\Gamma}\xspace}

\begin{document}

\begin{frontmatter}

\title{Parameters of charmonium states from KEDR}

\author[binp,nsu]{V.\,M.\,Aulchenko}
\author[binp,nsu]{E.\,M.\,Baldin\corref{cor}}
\cortext[cor]{Corresponding author, e-mail:  E.M.Baldin@inp.nsk.su}
\author[binp]{A.\,K.\,Barladyan}
\author[binp]{A.\,Yu.\,Barnyakov}
\author[binp]{M.\,Yu.\,Barnyakov}
\author[binp,nsu]{S.\,E.\,Baru}
\author[binp]{I.\,Yu.\,Basok}
\author[binp]{A.\,M.\,Batrakov} 
\author[binp]{A.\,E.\,Blinov}
\author[binp,nstu]{V.\,E.\,Blinov}
\author[binp]{A.\,V.\,Bobrov}
\author[binp]{V.\,S.\,Bobrovnikov}
\author[binp,nsu]{A.\,V.\,Bogomyagkov}
\author[binp,nsu]{A.\,E.\,Bondar}
\author[binp]{A.\,R.\,Buzykaev}
\author[binp,nsu]{S.\,I.\,Eidelman}
\author[binp,nsu,nstu]{D.\,N.\,Grigoriev}
\author[binp]{V.\,R.\,Groshev} 
\author[binp]{Yu.\,M.\,Glukhovchenko}
\author[binp]{V.\,V.\,Gulevich}
\author[binp]{D.\,V.\,Gusev}
\author[binp]{S.\,E.\,Karnaev}
\author[binp]{G.\,V.\,Karpov}
\author[binp]{S.\,V.\,Karpov}
\author[binp]{T.\,A.\,Kharlamova}
\author[binp]{V.\,A.\,Kiselev}
\author[binp]{V.\,V.\,Kolmogorov}
\author[binp,nsu]{S.\,A.\,Kononov}
\author[binp]{K.\,Yu.\,Kotov}
\author[binp,nsu]{E.\,A.\,Kravchenko}
\author[binp]{V.\,N.\,Kudryavtsev}
\author[binp,nsu]{V.\,F.\,Kulikov}
\author[binp,nstu]{G.\,Ya.\,Kurkin}
\author[binp,nsu]{E.\,A.\,Kuper}
\author[binp]{I.\,A.\,Kuyanov}
\author[binp,nstu]{E.\,B.\,Levichev}
\author[binp,nsu]{D.\,A.\,Maksimov}
\author[binp]{V.\,M.\,Malyshev}
\author[binp]{A.\,L.\,Maslennikov}
\author[binp,nsu]{O.\,I.\,Meshkov}
\author[binp]{S.\,I.\,Mishnev}
\author[binp,nsu]{I.\,I.\,Morozov}
\author[binp,nsu]{N.\,Yu.\,Muchnoi}
\author[binp]{V.\,V.\,Neufeld}
\author[binp]{S.\,A.\,Nikitin}
\author[binp,nsu]{I.\,B.\,Nikolaev}
\author[binp]{I.\,N.\,Okunev}
\author[binp,nstu]{A.\,P.\,Onuchin}
\author[binp]{S.\,B.\,Oreshkin}
\author[binp,nsu]{I.\,O.\,Orlov}
\author[binp]{A.\,A.\,Osipov}
\author[binp]{I.\,V.\,Ovtin}
\author[binp]{S.\,V.\,Peleganchuk}
\author[binp,nstu]{S.\,G.\,Pivovarov}
\author[binp]{P.\,A.\,Piminov}
\author[binp]{V.\,V.\,Petrov}
\author[binp]{A.\,O.\,Poluektov}
\author[binp]{V.\,G.\,Prisekin}
\author[binp,nsu]{O.\,L.\,Rezanova}
\author[binp]{A.\,A.\,Ruban}
\author[binp]{V.\,K.\,Sandyrev}
\author[binp]{G.\,A.\,Savinov}
\author[binp]{A.\,G.\,Shamov}
\author[binp]{D.\,N.\,Shatilov}
\author[binp,nsu]{B.\,A.\,Shwartz}
\author[binp]{E.\,A.\,Simonov}
\author[binp]{S.\,V.\,Sinyatkin}
\author[binp]{A.\,N.\,Skrinsky}
\author[binp,nsu]{A.\,V.\,Sokolov}
\author[binp]{A.\,M.\,Sukharev}
\author[binp,nsu]{E.\,V.\,Starostina}
\author[binp,nsu]{A.\,A.\,Talyshev}
\author[binp]{V.\,A.\,Tayursky}
\author[binp,nsu]{V.\,I.\,Telnov}
\author[binp,nsu]{Yu.\,A.\,Tikhonov}
\author[binp,nsu]{K.\,Yu.\,Todyshev}
\author[binp]{G.\,M.\,Tumaikin}
\author[binp]{Yu.\,V.\,Usov}
\author[binp]{A.\,I.\,Vorobiov}
\author[binp]{V.\,N.\,Zhilich}
\author[binp,nsu]{V.\,V.\,Zhulanov}
\author[binp,nsu]{A.\,N.\,Zhuravlev}

\address[binp]{Budker Institute of Nuclear Physics, 11, akademika
  Lavrentieva prospect,  Novosibirsk, 630090, Russia}
\address[nsu]{Novosibirsk State University, 2, Pirogova street,  Novosibirsk, 630090, Russia}
\address[nstu]{Novosibirsk State Technical University, 20, Karl Marx
  prospect,  Novosibirsk, 630092, Russia}

\begin{abstract}

  We report results of experiments performed with the KEDR detector at
  the \mbox{VEPP-4M} $e^+e^-$ collider. They include final results for the
  mass and other parameters of the
  \(J/\psi\), \(\psi(2S)\) and \(\psi(3770)\) and
  $J/\psi\to\gamma\eta_c$ 
  branching fraction determination.
\end{abstract}
\end{frontmatter}

\section{VEPP-4M collider}
\label{sec:VEPP}

The VEPP-4M collider~\cite{Anashin:1998sj} with the KEDR
detector~\cite{KEDR-detector} can operate in the wide range of beam
energy from 1 to 6 GeV.  The luminosity is quite modest, but
VEPP-4M is equipped with two independent systems for precise energy
determination~\cite{Blinov:2009zza}: the resonant depolarization facility
and infrared light Compton backscattering one.


\section{$J/\psi$ and $\psi(2S)$ mass measurement}
\label{sec:jpsi-psi2s-masses}

\begin{figure}[t]
  \centering
  \includegraphics[width=0.49\textwidth]{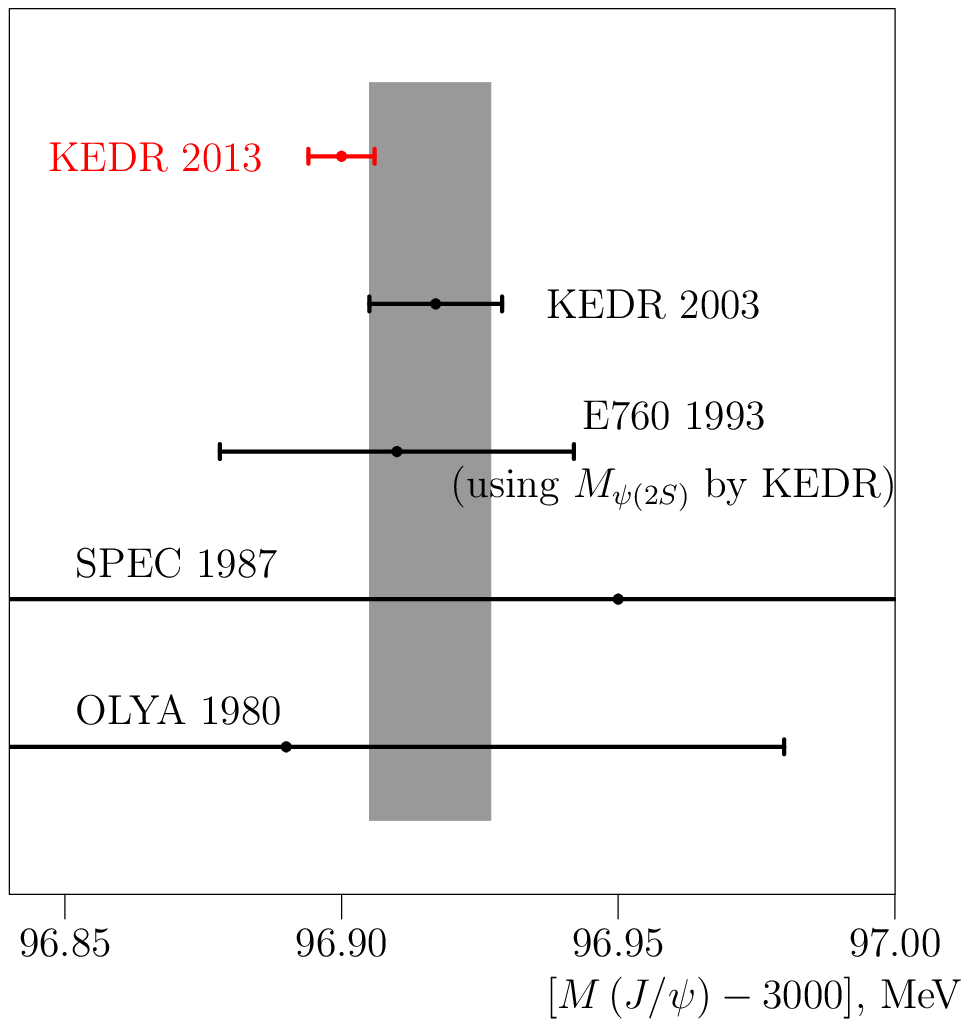}\hfill
  \includegraphics[width=0.49\textwidth]{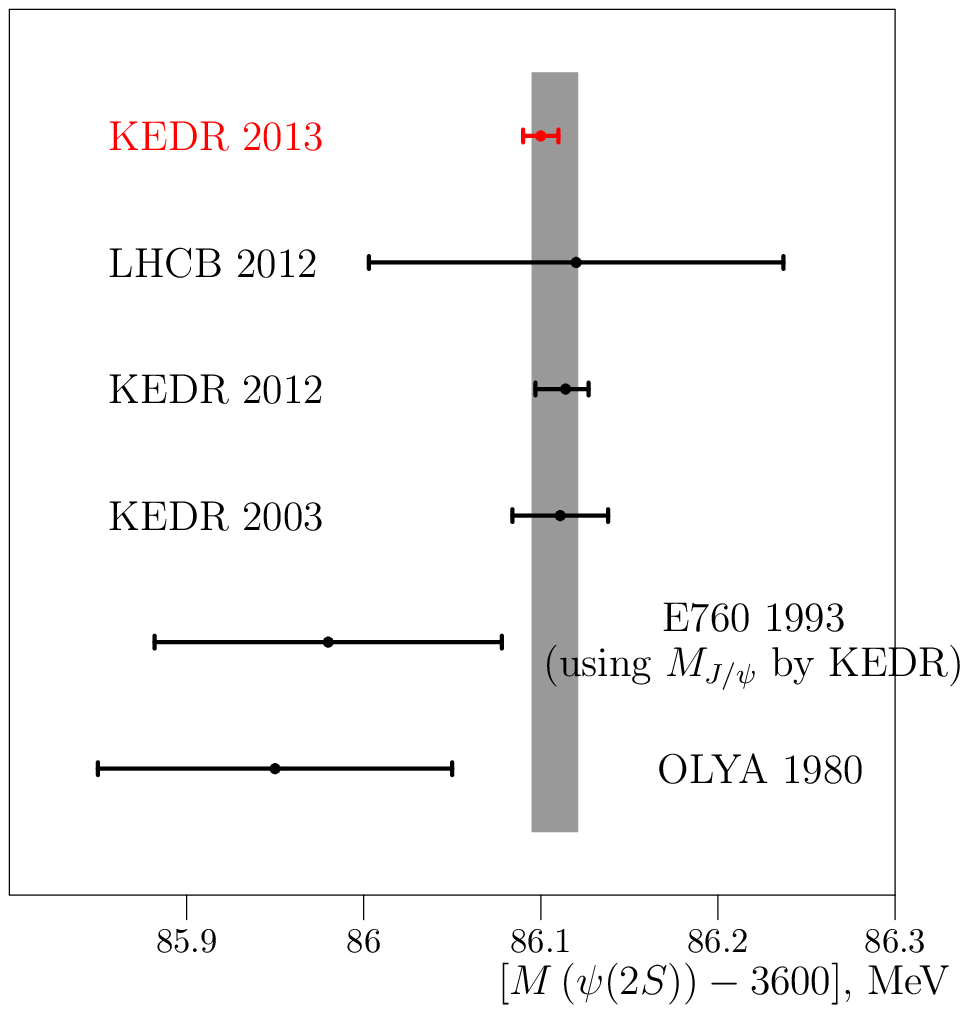}
  \parbox{0.49\textwidth}{\caption{$J/\psi$ mass comparison.}\label{fig:Mjpsi}}
  \parbox{0.49\textwidth}{\caption{$\psi(2S)$ mass comparison.}\label{fig:Mpsi2S}}
\end{figure}

The  $J/\psi$ and $\psi(2S)$  masses were
measured using additional statistics. Systematic
errors in mass measurements are the main issue. More than 20 different
effects were considered: energy spread, energy assignment, energy
difference of $e^+$ and $e^-$, beam misalignment, luminosity etc.  New
mass measurements will supersede the old results~\cite{Aulchenko:2003qq}.

The final mass values obtained are:  
\[
\begin{split}
  M\left(J/\psi\right)&=(3096.900 \pm 0.002 \pm 0.006)\,\text{MeV,}\\
  M\left(\psi(2S)\right)&=(3686.100 \pm 0.004 \pm 0.009)\,\text{MeV.}
\end{split}
\]

A comparison with other measurements is presented in
Figs.~\ref{fig:Mjpsi} and \ref{fig:Mpsi2S} for the $J/\psi$ and
$\psi(2S)$ masses, respectively. The position and width of the bar
correspond to the PDG2012 fit~\cite{PDG-2012}.

\section{Comparison of the $J/\psi$ leptonic widths}
\label{sec:GeeVsGmumu}

\begin{figure}[t]
  \centering
  \includegraphics[width=0.50\textwidth]{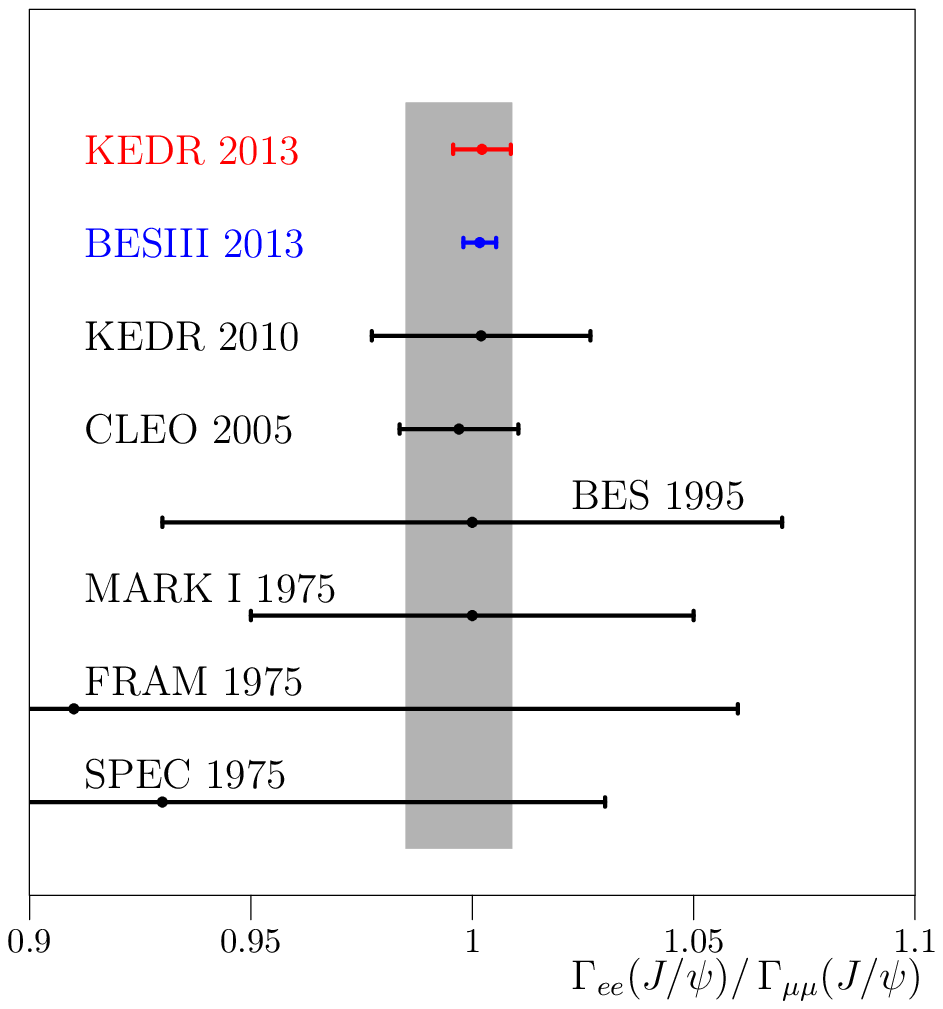}\hfill
  \includegraphics[width=0.48\textwidth]{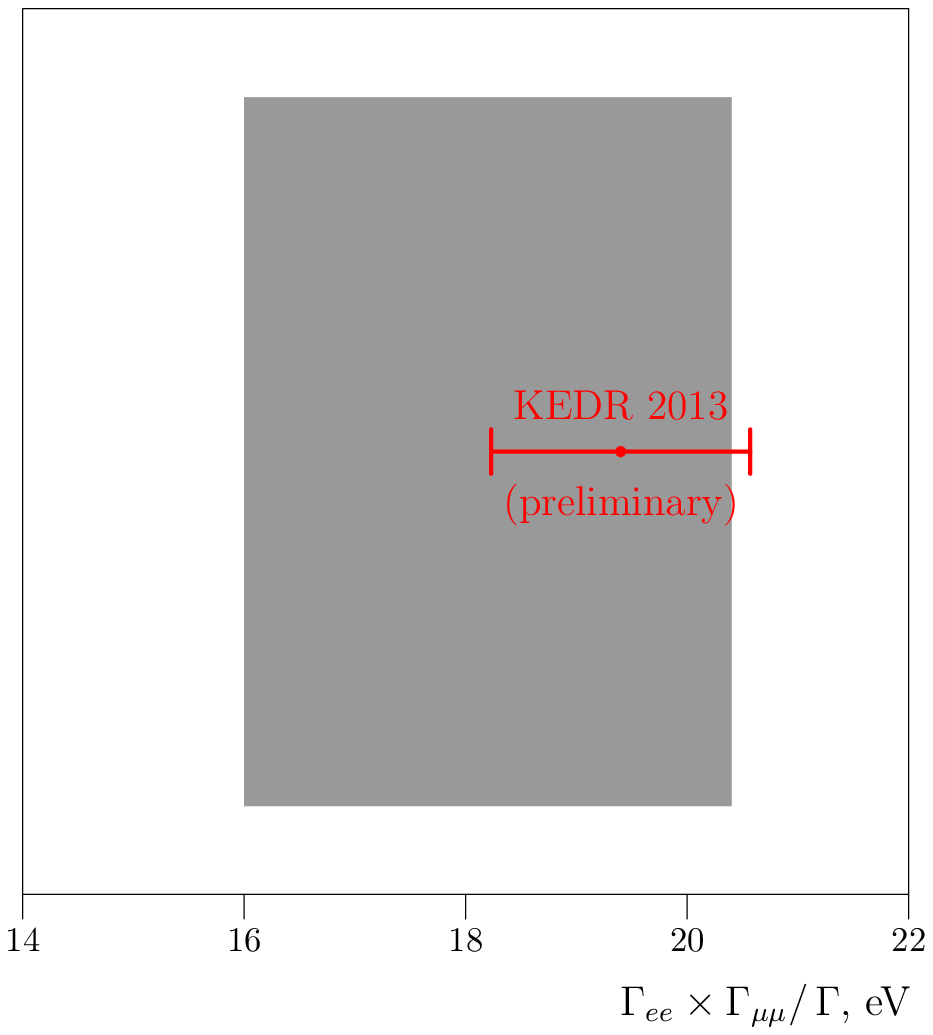}
  \parbox{0.49\textwidth}{\caption{$J/\psi$ $\Gamma_{ee}/\Gamma_{\mu\mu}$
  comparison.}\label{fig:GeeVsGmumu}}
  \parbox{0.49\textwidth}{\caption{$\psi(2S)$ \GBmumu comparison.}
    \label{fig:GBmumupsi2S}}
\end{figure}

The ratio of the electron and muon widths of the \(J/\psi\) meson has
been measured at the KEDR detector using direct $J/\psi$ decays.  The result
\[\Gamma_{e^+e^-}(J/\psi)/\Gamma_{\mu^+\mu^-}(J/\psi)=1.0022\pm0.0044\pm0.0048\ (0.65\%)\]
is in good agreement with the lepton universality. The experience
collected during this analysis will be used for a \(J/\psi\) lepton
width determination with up to 1\% accuracy.

A comparison with other measurements is presented in
Fig.~\ref{fig:GeeVsGmumu}. The position and width of the bar
correspond to the PDG2012 fit~\cite{PDG-2012}. Currently the world
average value of the \(J/\psi\) meson lepton width is completely
dominated by the CLEO results obtained in 2005~\cite{Li:2005uga}.
Recently the BESIII collaboration announced the most precise
measurement of the ratio of the electron and meson
widths~\cite{Ablikim:2013pqa}.  For that analysis both experiments
used the \(\psi(2S)\to J/\psi\pi^+\pi^-\), \(J/\psi\to\ell^+\ell^-\)
decay chain (\(\ell=e,\mu\)).

\section{Study of $\psi(2S)\to\mumu$ decay}
Since 2004 KEDR has been taking data at the VEPP-4M collider   in 
the $\psi(2S)$ region and acquired a total luminosity of about 7 pb$^{-1}$, 
which corresponds to more than $3.5 \times 10^6$ $\psi(2S)$.

We report the preliminary value of
\[\Gamma_{ee}(\psi(2S))\times\Gamma_{\mu\mu}(\psi(2S))/\Gamma(\psi(2S))=(19.4\pm0.4\pm1.1)\,\text{eV}.\]

No direct measurement of this quantity is listed in the PDG tables. For
comparison (the position and width of the bar in
Fig.~\ref{fig:GBmumupsi2S}) we use the product of the two world average
values -- \Gee and $\mathcal{B}_{\mu\mu}$.

\section{The main parameters of the $\psi(3770)$}

\begin{figure}[t]
  \centering
  \includegraphics[width=0.50\textwidth]{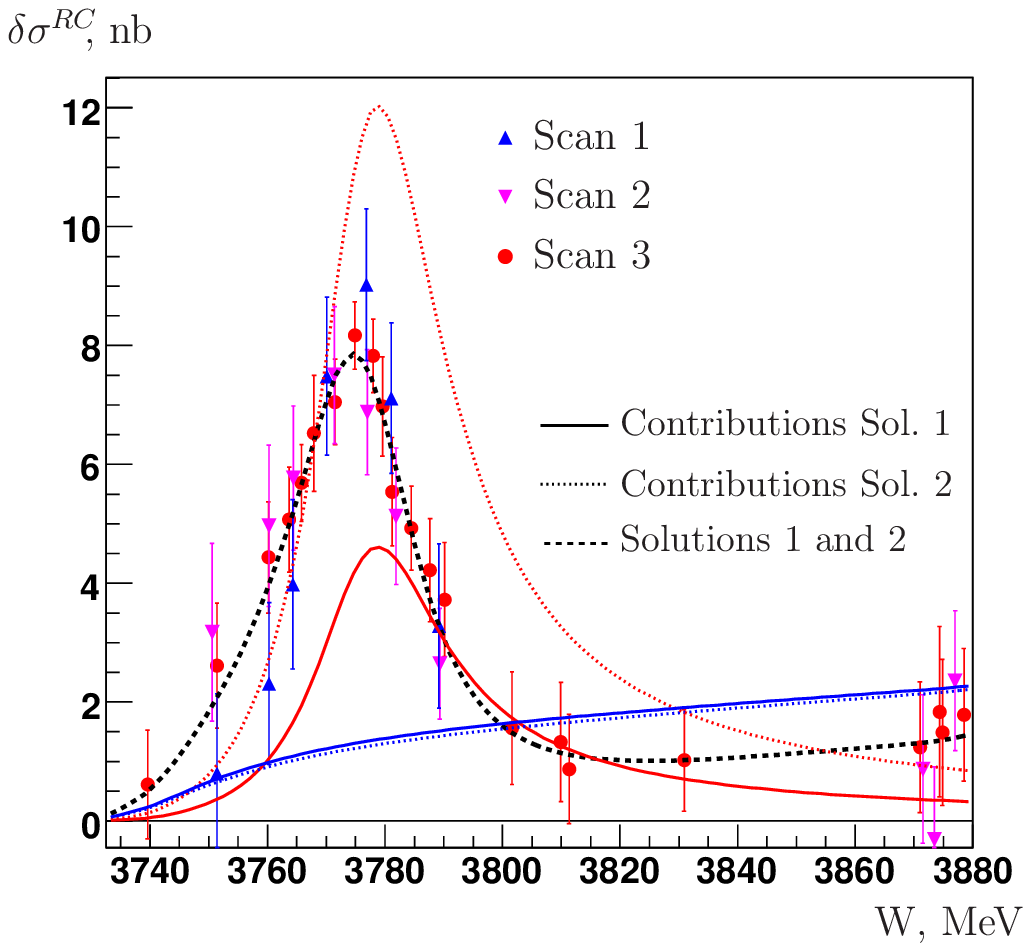}\hfill
  \includegraphics[width=0.45\textwidth]{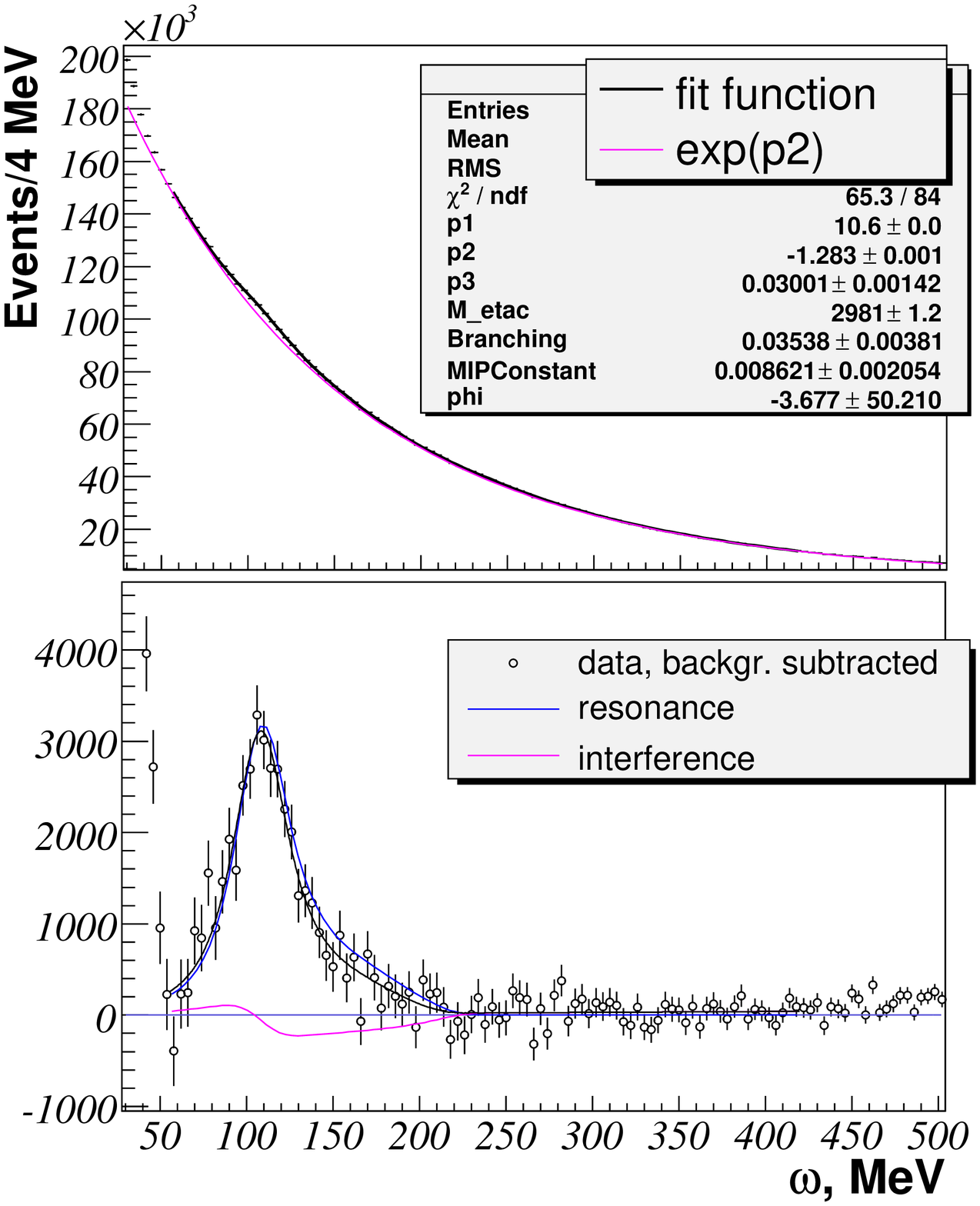}
  \parbox{0.50\textwidth}{\caption{The \(\psi(3770)\) cross section
      with two fit solutions.}\label{fig:twofitdp2012}}\hfill
  \parbox{0.45\textwidth}{\caption{The inclusive photon spectrum and
      the \(\eta_c\) observed cross
      section.}\label{fig:FitGNfixedHint2}}
\end{figure}

For the \(\psi(3770)\) mass measurement the joint \(\psi(2S)\) and
\(\psi(3770)\) scans were performed. Data analysis takes into account
interference between the resonant and nonresonant $\DD$ production.

Our final results for mass and width are~\cite{Anashin:2011kq}:
\begin{equation*}
  \begin{split}
    M(\psi(3770)) & =  3779.2\,\,^{+1.8}_{-1.7}\,\, ^{+0.5}_{-0.7} \,\,^{+0.3}_{-0.3} \,\, \text{MeV},  \\
    \Gamma(\psi(3770)) & = \:\:\:\:\,24.9\,\,^{+4.6}_{-4.0}\,\, ^{+0.5}_{-0.6}\,\,^{+0.2}_{-0.9} \,\,    \text{MeV}.
  \end{split}
\end{equation*}

When a resonance interferes with a variable continuum,
the ambiguity appears. Figure~\ref{fig:twofitdp2012} 
illustrates two solutions corresponding to VDM.  The resulting
cross sections, mass and width are almost indistinguishable, but for
the lepton width and phase we have two solutions:
\begin{equation*}
  \begin{split}
    &(1) \:\:~\Gamma_{ee}(\psi(3770)) =
    154\,^{+79}_{-58}\,^{+17}_{-9}\,^{+13}_{-25}\: \text{eV}, 
    \quad \phi=(171\pm17)\,\grad,\\
    &(2) \:\:~\Gamma_{ee}(\psi(3770))=
    414\,^{+72}_{-80}\,^{+24}_{-26}\,^{+90}_{-10}\: \text{eV},
    \quad \phi=(240\pm9)\,\grad.
  \end{split}
\end{equation*}

Recently the VDM approach was applied for the joint analysis of BABAR,
BELLE, BES, CLEO and KEDR data on $D\overline{D}$-- and inclusive
hadronic cross sections~\cite{psi3770jointfit-poster2013}:\par

\begin{center}
  \begin{tabular}{cccccc}\hline
    Solution&  $M$ (MeV) & $\Gamma$ (MeV) & $\mathcal{B}_{nD\bar{D}}$ & $\Gamma_{ee}$(eV) &  $\phi$ (deg)  \\*[0.5ex]
    \hline
    1 & 3779.3$\pm$1.0 & 26.7$\pm$1.4 & 0.19$\pm$0.05 & 202$\pm$18 & 185.6$\pm$5.2\\
    2 & 3779.5$\pm$1.0 & 26.8$\pm$1.4 & 0.11$\pm$0.03 & 346$\pm$19 & 228.6$\pm$3.0\\\hline
  \end{tabular}
\end{center}
The results of KEDR were confirmed at the much better level of
accuracy.

\section{$\mathcal{B}(J/\psi\to\gamma\eta_c)$}

The $\eta_c$ mass, width and branching fraction of
$J/\psi\to\gamma\eta_c$ decay
have been measured in the inclusive photon spectrum of multihadron
$J/\psi$ decays (Fig.~\ref{fig:FitGNfixedHint2}), using a sample of
about 6 million $J/\psi$ mesons.

$J/\psi\to\gamma\eta_c$ decay is an M1 transition between the 1S
states of charmonium, so its rate can be easily calculated in
potential models in the limit of a zero width of the resonance. For
this decay, the resonance width to transition energy ratio is about $1/4$,
thus the photon line shape deviates from the Breit-Wigner.
During the analysis the \(\omega^3\)  factor near the $\eta_c$ resonance
and interference effects were taken into account.

Our final result for the branching fraction of $J/\psi\to\gamma\eta_c$
is:
\[\mathcal{B}(J/\psi\to\gamma\eta_c) = (3.58 \pm 0.38 \pm 0.20)\%\]

\section{Summary}

Most of the results are in good agreement with the world average
values if they exist and have comparable or better accuracy. 

The results of the KEDR detector on $\psi(3770)$ are confirmed by joint
analysis of data published by 5 experiments at a better accuracy level.

This work is supported by the Ministry of Education and Science of the
Russian Federation, the RFBR grants \mbox{12-02-00023}, 11-02-00112,
11-02-00558, the Sci. School Nsh-5320.2012.2 grant and the DFG grant HA
1457/7-2.


\end{document}